# Confidence intervals for adaptive trial designs II: Case study and practical guidance

**David S. Robertson**[1*], **Thomas Burnett**[2], **Babak Choodari-Oskooei**[3], **Munya Dimairo**[4], **Michael Grayling**[5], **Philip Pallmann**[6], **Thomas Jaki**[1,7]

[1] *MRC Biostatistics Unit, University of Cambridge, UK*
[2] *University of Bath, UK*
[3] *MRC Clinical Trials Unit at UCL, UK*
[4] *School of Health and Related Research (ScHARR), University of Sheffield, UK*
[5] *Johnson & Johnson Innovative Medicine*
[6] *Centre for Trials Research, Cardiff University, UK*
[7] *University of Regensburg, Germany*
[*] *Corresponding author:* david.robertson@mrc-bsu.cam.ac.uk

## Abstract

In adaptive clinical trials, the conventional confidence interval (CI) for a treatment effect is prone to undesirable properties such as undercoverage and potential inconsistency with the final hypothesis testing decision. Accordingly, as is stated in recent regulatory guidance on adaptive designs, there is the need for caution in the interpretation of CIs constructed during and after an adaptive clinical trial. However, it may be unclear which of the available CIs in the literature are preferable. This paper is the second in a two-part series that explores CIs for adaptive trials. Part I provided a methodological review of approaches to construct CIs for adaptive designs. In this paper (part II), we present an extended case study based around a two-stage group sequential trial, including a comprehensive simulation study of the proposed CIs for this setting. This facilitates an expanded description of considerations around what makes for an effective CI procedure following an adaptive trial. We show that the CIs can have notably different properties. Finally, we propose a set of guidelines for researchers around the choice of CIs and the reporting of CIs following an adaptive design.

## 1. Introduction

Clinical trials are traditionally run in a fixed manner that does not allow for interim looks at the data within the trial itself. In contrast, an adaptive design (AD) allows for pre-planned modifications to the course of the trial based on interim data analyses[1–3]. This added flexibility can lead to improved trial efficiency (e.g., in terms of sample size, time and cost) while still maintaining scientific rigour. ADs have seen increasing use in clinical trials in recent years, and in particular master protocols leveraging ADs are becoming increasingly popular[4,5].

A wide variety of different types of AD have been proposed in the literature, including the following common broad classes:

- Early trial stopping: *Group sequential designs (GSD)* allow the trial to stop early at interim looks for efficacy or futility/lack-of benefit.

- Treatment selection: *Multi-arm multi-stage (MAMS)* designs test multiple treatment options in parallel (typically against a common control arm), allowing the dropping of treatment arm(s) that are not performing (as) well.

- Population selection: *Adaptive enrichment* designs allow the clinical (sub)population of interest to be selected ('enriched') at interim looks, typically using pre-defined patient subpopulations based on biomarker information.

- Changing randomisation probabilities: *Response-adaptive randomisation (RAR)* allows updates of the randomisation probabilities based on patient responses, for example to favour treatment arm(s) that are performing well[6].

- Changing trial sample size: *Sample size re-estimation* allows the sample size of the trial to be adjusted, for example based on interim conditional power calculations.

Further educational material on all of these ADs can be found in Burnett et al. (2020)[7], Pallmann et al. (2018)[1], and the PANDA online resource (https://panda.shef.ac.uk/)[8].

Regardless of the type of AD, it remains crucial that the integrity and validity of the trial is maintained. Appropriate estimation of treatment effects is a key part of trial validity, which includes not only point estimates (see Robertson et al. 2023a,b[9,10]) but also quantification of the uncertainty around the estimated treatment effects as given by *confidence intervals* (CIs). Intuitively, CIs capture this uncertainty by offering an interval that is expected to typically contain the unknown parameter of interest.

An important consideration in practice then is whether proposed methods to construct CIs have desirable properties. Most importantly, this relates to the CI having the desired coverage probability (i.e., the long-run probability that the CI contains the true unknown treatment effect of interest). However, there are numerous other considerations, including the width of the CI (all else being equal, narrower CIs are preferred as they are more informative), whether the CI will always contain an associated point estimate, and whether the CI will always be consistent with the decision rule (i.e., with an associated hypothesis test). The fundamental problem for ADs is that use of standard CI methodology (i.e., CIs constructed using methods that do not account for the fact an AD has been used) may not necessarily result in desirable properties.

Recent regulatory guidance highlights these concerns, with the U.S. Food and Drug Administration (FDA) noting that "confidence intervals for the primary and secondary endpoints may not have correct coverage probabilities for the true treatment effects" and thus "confidence intervals should be presented with appropriate cautions regarding their interpretation"[11]. The same guidance also highlights the need to pre-specify methods used to compute CIs after an AD, see also the Adaptive designs CONSORT Extension (ACE) guidance[2,3]. Meanwhile, the European Medicines Agency (EMA) guidance states that "methods to … provide confidence intervals with pre-specified coverage probability are required" if an AD is used in a regulated setting[12].

These concerns and regulatory guidance motivate the growing body of literature proposing 'adjusted' CIs that are specifically tailored for use with a particular AD. However, in our experience there is at best limited uptake of adjusted CIs in practice (with the possible exception of 'repeated' CIs in GSDs; see part I of the paper series and Section 3.1 of this current paper for a definition), with many adaptive trials continuing to only report the standard CI. Evidence for this in the context of two-stage single-arm trials can be found in Grayling and Mander (2021)[13], where only 2% of 425 articles reported an adjusted CI. This limited uptake of adjusted CIs for ADs is due to a number of reasons, including the lack of awareness of methods in the literature, available software/code and guidance around the choice of adjusted CI in practice.

This paper is the second in a two-part series that explores the issue of CIs for ADs. In part I of the series, we reviewed and compared methods for constructing CIs for different classes of ADs and critically discussed different approaches. In the current paper (part II), we consider CIs for ADs from a practical perspective, and propose a set of guidelines for researchers around the choice and reporting of CIs following an AD. We first briefly describe performance measures of CIs in Section 2. We introduce the case study in Section 3 and show how to calculate different types of CIs (with R code provided). We use the case study as a basis for a simulation study in Section 4. We conclude with guidance for researchers and discussion in Sections 5 and 6, respectively.

## 2. Performance measures for CIs

As introduced in part I of this paper series, different desirable properties for CIs have been proposed, which we recapitulate below. To fix ideas, suppose we have a random sample $X$ from a probability distribution with parameter $\theta$, which is the single parameter of interest in the trial. A CI for $\theta$ with confidence level $1 - \alpha$ is a random interval $(L(X), U(X))$ that has the following (claimed) property: $P(L(X) < \theta < U(X)) = 1 - \alpha$ for all $\theta$.

The *coverage* probability (often shortened to just 'coverage') of a confidence interval $(L(X), U(X))$ is given by $P(L(X) < \theta < U(X))$. This is the key performance measure for a CI, given that the definition itself of a CI is based around actually having the claimed 'nominal' $1 - \alpha$ confidence level.

Alongside coverage, other criteria for CIs (generally, and specifically for ADs) have been proposed in the literature. The main criteria/performance measures include:

- Correct coverage (arguably essential)
- Width (all other things being equal, a smaller width is desirable)
- Consistency/compatibility with the hypothesis test (see below)
- Contains the point estimate of interest
- Is in fact an interval (i.e., not a union of disjoint intervals, or the empty set)
- Is computationally feasible/simple

A CI is *consistent/compatible* with the hypothesis testing decision if it excludes the parameter value(s) that are rejected by the hypothesis test, and conversely excludes the parameter value(s) that are *not* rejected by the hypothesis test. If a CI is *not* consistent/compatible with the hypothesis testing decision then this can lead to problems with study interpretation and the communication of results.

## 3. Case study

As an example, we use the phase III MUSEC (multiple sclerosis and extract of cannabis) group sequential trial (Bauer et al., 2016[14]; Zajicek et al., 2012[15]). MUSEC investigated a standardised oral cannabis extract (CE), assessing its effect on muscle stiffness in adults with stable multiple sclerosis compared to placebo. Its primary outcome was whether or not a patient had "relief from muscle stiffness" after 12 weeks of treatments, based on a dichotomised 11-point category rating scale. MUSEC utilised a two-stage GSD, with early stopping for superiority assessed using an O'Brien-Fleming (OBF) type boundary. The unblinded interim analysis was planned after 200 participants (100 per arm) had completed the 12 week treatment course, with the final analysis planned after 400 patients (200 per arm) if the interim stopping rule was not met. The actual trial did plan for sample size re-estimation too, but for simplicity we will focus on the group sequential aspect here.

Ultimately, the trial did continue to its second stage; Table 1 summarises the study data at the interim and final analyses, as well as the OBF efficacy stopping boundaries. As can be seen, at the interim analysis the boundary for early rejection of the null hypothesis (no difference in the proportion of subjects with relief from muscle stiffness between treatment arms) was almost crossed, with the standardised test statistic being close to the stopping boundary.

|  | Interim analysis | | Final analysis | |
| --- | --- | --- | --- | --- |
|  | Placebo | Cannabis extract | Placebo | Cannabis extract |
| Number of patients with relief from muscle stiffness | 12 | 27 | 21 | 42 |
| Total number of subjects | 97 | 101 | 134 | 143 |
| Standardised test statistic | 2.540 | | 2.718 | |
| O'Brien-Fleming stopping boundary | 2.797 | | 1.977 | |

*Table 1: MUSEC trial observed data, by analysis stage. The O'Brien-Fleming efficacy stopping boundaries are also shown.*

Typically, at the final analysis a $100(1-\alpha)$% CI will be desired for the difference in the response rates in the placebo and CE arms (the measure of the treatment effect of interest). A common method of achieving this is to use a standard two-sided interval, e.g., based on Wald's methodology[16]. For MUSEC, denoting the final sample sizes in the two arms by $n_P = 134$ and $n_{CE} = 143$, and the MLEs for the rates by $\hat{p}_P = 21/n_P$ and $\hat{p}_{CE} = 42/n_{CE}$, this gives for $\alpha = 0.05$

$$(\hat{p}_{CE} - \hat{p}_P) \pm \Phi^{-1}(1-\alpha/2)\widehat{Var}(\hat{p}_{CE} - \hat{p}_P) = (\hat{p}_{CE} - \hat{p}_P) \pm \Phi^{-1}(1-\alpha/2)\sqrt{\frac{\hat{p}_{CE}(1-\hat{p}_{CE})}{n_{CE}} + \frac{\hat{p}_P(1-\hat{p}_P)}{n_P}}$$
$$= (0.040, 0.234),$$

where $\Phi^{-1}(x)$ denotes the inverse cumulative distribution function (CDF) of a standard normal random variable.

If the interim analysis was not present, this CI would have a number of desirable properties (as discussed in Section 2): it would be guaranteed to be an interval containing the MLE for $\hat{p}_{CE} - \hat{p}_P$, would (at least asymptotically) have the desired coverage, would be consistent with the associated hypothesis test, and would evidently be easily computed.

In this Section (as well as the simulation study in Section 4), we illustrate how different types of CIs (both unconditional and conditional) can be used in practice for a GSD, based on the MUSEC trial. We use a GSD as our case study in order to illustrate the widest range of different CIs.

## 3.1 Calculation of CIs when continuing to stage 2

Using the observed data from the MUSEC trial, we now show how to calculate various CIs for the treatment difference, denoted $\theta = p_{CE} - p_P$, from both a conditional and unconditional perspective (see explanation on the difference below) when the trial continues to stage 2, as happened for the MUSEC trial. R code to obtain all CIs considered is provided in the data files.

As already seen above, the **standard/naive CI** for the treatment difference (i.e., the Wald CI), is given by

$$\hat{\theta} \pm \Phi^{-1}(1-\alpha/2)\widehat{Var}(\hat{\theta}) = (\hat{p}_{CE} - \hat{p}_P) \pm \Phi^{-1}(1-\alpha/2)\sqrt{\hat{p}_{CE}(1-\hat{p}_{CE})/n_{CE} + \hat{p}_P(1-\hat{p}_P)/n_p}$$

Other methods than Wald are available to calculate standard/naive CIs for the difference of two proportions[17,18], but none of these will be able to account for the AD used.

From an unconditional perspective, we want to estimate θ regardless of the stage that the trial stops, and are interested in the properties of the CI as averaged over all possible stopping times (weighted by the respective stage-wise stopping probabilities). Note that the standard/naive CI above is an unconditional CI.

In what follows, we let $e_1$, $e_2$ denote the efficacy stopping boundaries, $I_1$, $I_2$ the (observed) information, at stages 1 and stage 2, respectively, and $T$ the stage the trial stopped at (so $T = 2$ for the MUSEC trial). The definitions of the information $I_1$ and $I_2$ for the MUSEC trial are given in Appendix A.1, which depend on the number of observed successes.

The **exact unconditional CI** depends on a choice of the ordering of the sample space with respect to evidence against the null hypothesis. In what follows, we use stagewise ordering, which has desirable properties described by Jennison and Turnbull (1999)[19]. This allows the use of the *p*-value function P(θ) to find the lower and upper bounds for the 95% CI, $\hat{\theta}_l$ and $\hat{\theta}_u$, which are the solutions to the equations $P(\hat{\theta}_l) = 0.025$ and $P(\hat{\theta}_u) = 0.975$. The formula for the *p*-value function for stopping stage $T = 2$ (with observed second stage test statistic $Z_2 = z_2$) is as follows:

$$P(\theta) = \int_{-\infty}^{e_1} \int_{z_2}^{\infty} f_2\left((x_1, x_2), (\theta\sqrt{I_1}, \theta\sqrt{I_2}), \begin{pmatrix} 1 & \sqrt{I_1/I_2} \\ \sqrt{I_1/I_2} & 1 \end{pmatrix}\right) dx_2\, dx_1$$

where $f_2((\boldsymbol{x_1}, \boldsymbol{x_2}), \mu, \Sigma)$ is the density of a bivariate normal distribution with mean vector $\mu$ and covariance matrix $\Sigma$ evaluated at the vector $(\boldsymbol{x_1}, \boldsymbol{x_2})$. Note that the associated point estimator of this method, the median unbiased estimator (MUE), $\hat{\theta}_{MUE}$, is the solution to the equation $P(\hat{\theta}_{MUE}) = 0.5$. If the distributional assumptions hold exactly (i.e., the joint canonical distribution of test statistics holds exactly), then the exact unconditional CI guarantees consistency with the test decision.

The **repeated CI (RCI)** follows a simple form: $\hat{\theta} \pm e_T / \sqrt{I_T}$, see Jennison and Turnbull (1999)[19]. Note that there is no explicit associated point estimator with this method (although one could of course just use the standard MLE $\hat{\theta}$). Since $\hat{\theta} = Z_T / \sqrt{I_T}$, the RCI guarantees consistency with the test decision.

The **adjusted asymptotic CI** adjusts the standard CI, giving a CI of the form

$$(\hat{\theta} - \hat{\mu}(\hat{\theta})) \pm \Phi^{-1}(1 - \alpha/2)\hat{\sigma}(\hat{\theta})/\sqrt{I_T}$$

where $\hat{\mu}(\hat{\theta})$ and $\hat{\sigma}(\hat{\theta})$ are functions of $\hat{\theta}$ given in Todd et al. (1996)[20]:
$\hat{\mu}(\hat{\theta}) = E(\hat{\theta}) - \hat{\theta} E(\sqrt{I_T})$
$\hat{\sigma}(\hat{\theta}) = E(Z_T^2/I_T) - 2\hat{\theta} E(Z_T) + \hat{\theta}^2 E(I_T) - \hat{\mu}(\hat{\theta})^2$.

The associated point estimator for this method is the bias-adjusted estimator $\hat{\theta} - \hat{\mu}(\hat{\theta})$.

One subtlety with the use of the adjusted asymptotic CI (for trials that continue to stage 2) is that it is possible for the information levels to actually *decrease* from stage 1 to stage 2 i.e., $I_2 < I_1$. This can happen in trials with a binary outcome when the pooled estimated response rate is very close to zero (or 1) at stage 1, but further away from zero (and 1) at stage 2. In this situation, it is not possible to calculate the adjusted asymptotic CI, although this happens very rarely (i.e., one or two times out of $10^5$ simulation replicates) for trials with success rates similar to those observed in the MUSEC trial. It is similarly possible for this to occur in trials with, e.g., normal (if the estimated standard deviation differs greatly between stages) and time-to-event data (typically when the number of events that occur between analyses is smaller), though again is a rare occurrence.

For the **parametric bootstrap CI** we use the a simple bootstrap algorithm to generate *B* bootstrap MLEs $\hat{\theta}^{(b)}$ for $b = 1, \ldots, B$. In the interests of space, we defer the details to Appendix A.1. Note that the associated point estimator for this method is given by the mean of $\hat{\theta}^{(1)}, \ldots, \hat{\theta}^{(B)}$.

Finally, the **randomisation-based CI** follows a different bootstrap procedure in order to calculate an adjusted p-value, based on the randomisation distribution (this being an example of randomisation-based inference i.e., a randomisation test). The idea is to reproduce the group sequential analysis for each allowable allocation of patients to treatments, but *keeping the observed patient outcomes fixed*. The adjusted p-value is then the proportion of these potential results that are as extreme or more extreme than the actual trial result. Note that at the interim analysis, exactly the same set of patients are used each time (although with different treatment allocations).

More precisely, the procedure is defined as follows (Snapinn, 1994[21]). Suppose a group sequential trial has been completed, and let **X** denote the set of all observed patient outcomes.

Let the vector **T** denote the set of all possible allocations of patients to treatments (given the randomisation procedure used). Let $t_i$ denote a potential allocation of patients to treatments and $t^*$ denote the actual allocation used in the trial. Then let $F(\mathbf{X}, t_i)$ be the measure of strength of evidence against the null hypothesis. The adjusted p-value is then

$$p_{\text{adjusted}} = \sum_{t_i \in \mathbf{T}} I\{F(\mathbf{X}, t_i) \geq F(\mathbf{X}, t^*)\}/|\mathbf{T}|$$

For the measure of strength of evidence against the null hypothesis, we use stagewise ordering of the sample space (like for the unconditional exact CI). The randomisation-based CI is based on the adjusted p-value as follows:

$$\left[ \Phi^{-1}(1 - p_{adjusted}) \pm \Phi^{-1}(1 - \alpha/2) \right] \sqrt{\hat{p}_{CE}(1 - \hat{p}_{CE})/n_{CE} + \hat{p}_P(1 - \hat{p}_P)/n_p}$$

The associated point estimator for this CI is then

$$\Phi^{-1}(1 - p_{adjusted}) \sqrt{\hat{p}_{CE}(1 - \hat{p}_{CE})/n_{CE} + \hat{p}_P(1 - \hat{p}_P)/n_p}$$

In practice, it is not feasible (except for small sample sizes) to use the entire set of possible random allocations **T** and we use $N$ random samples from the set instead. However, a problem arises when $p_{adjusted} = 0$ i.e., no allocations are sampled that give results as extreme as or more extreme than those actually observed, which gives $\Phi^{-1}(1 - p_{adjusted}) = \infty$. Even when $N = 10{,}000$ this can occur as discussed later in the simulation study.

*Conditional CIs*

From a conditional perspective, we are interested in estimation conditional on the stage the trial stops at (so for the MUSEC trial, conditional on the trial continuing to stage 2).

The **exact conditional CI** uses the conditional density of the MLE, and is defined as:

$$CI_c = \{\theta : \alpha/2 < Prob(\hat{\theta} \geq \hat{\theta}_{obs} | T = t, \theta) < 1 - \alpha/2\}$$

where $\hat{\theta}_{obs}$ is the observed value of the MLE, see Fan and DeMets (2006)[22]. More explicitly, for a trial continuing to stage 2 (so $t = 2$) the lower and upper and bounds for the CI, $\hat{\theta}_l$ and $\hat{\theta}_u$, are the solutions to the equations

$$\alpha/2 = \int_{\hat{\theta}_{obs}}^{\infty} f(\hat{\theta} | \hat{\theta}_l, T = 2) d\hat{\theta} \quad \text{and} \quad 1 - \alpha/2 = \int_{\hat{\theta}_{obs}}^{\infty} f(\hat{\theta} | \hat{\theta}_u, T = 2) d\hat{\theta}$$

where $f(\hat{\theta} \mid \theta, T = 2)$ is the conditional density of the MLE (conditional on continuing to stage 2), see Appendix A.1 for further details. The associated point estimator is the conditional MUE, $\hat{\theta}_{CMUE}$, which is the solution to the following equation:

$$0.5 = \int_{\hat{\theta}_{obs}}^{\infty} f(\hat{\theta} \mid \hat{\theta}_{CMUE}, T = 2) d\hat{\theta}$$

Like for the adjusted asymptotic CI, if the information levels decrease from stage 1 to stage 2 i.e., $I_2 < I_1$, then the exact conditional CI (and hence the restricted exact conditional CI, see below) cannot be calculated.

The **restricted exact conditional CI**, as the name suggests, restricts the range of the exact conditional CI. Given the exact conditional CI ($CI_c$) defined above, the restricted exact conditional CI is defined to be $CI_c \cap CI_r$, where

$$CI_r = \{\theta : Prob(T \leq t \mid \theta) > \alpha/2 \text{ and } Prob(T \geq t \mid \theta) > \alpha/2\},$$

see Fan and DeMets (2006)[22]. For a trial continuing to stage 2 (so $t = 2$), the upper bound of the restricted exact conditional CI is set equal to $min\{\hat{\theta}_u, (e_1 - \Phi^{-1}(\alpha/2))/\sqrt{I_1}\}$, where $\hat{\theta}_u$ is the upper bound of $CI_c$.

The **conditional likelihood CI** is based on the conditional MLE i.e., the maximiser of the conditional log-likelihood. As shown in Marschner et al. (2022)[23], the conditional log-likelihood, conditioning on the trial stopping at stage $T = t$ is given by

$$L_c(\theta; z_t, t) = -\frac{1}{2}(z_t - \theta\sqrt{I_t})^2 - \log Pr(T = t \mid \theta).$$

Using the results of Fan and DeMets (2006)[22], we can show that the conditional MLE, $\hat{\theta}_c$, is the solution of the following equation when $T = 2$:

$$\hat{\theta}_c = \hat{\theta}_{obs} - \frac{\sqrt{I_1}\,\phi(e_1 - \hat{\theta}_c\sqrt{I_1})}{I_2\Phi(e_1 - \hat{\theta}_c\sqrt{I_1})}$$

where $\phi$ and $\Phi$ are the probability density function (pdf) and cdf of the standard normal distribution, respectively. The conditional MLE is the associated point estimator for the conditional likelihood CI. The conditional likelihood CI is calculated using a conditional bootstrap procedure, with full details provided in Appendix A.1.

The **penalised likelihood CI** is equal to the conditional likelihood CI when the trial continues to stage 2, see Section 3.2 for details of how it is different when in fact the trial stops at stage 1.

## 3.2 Calculation of CIs when stopping early

In this subsection we detail how the various CIs are calculated when the stopping stage is $T = 1$ (rather than $T = 2$ as in Section 3.1), which will be needed for the Simulation study in Section 4.

The **standard/naive CI** for the treatment difference (i.e., the Wald CI) is given by

$$\hat{\theta} \pm \Phi(1 - \alpha/2)\widehat{Var}(\hat{\theta}) = (\hat{p}_{CE} - \hat{p}_P) \pm \Phi(1 - \alpha/2)\sqrt{\frac{\hat{p}_{CE}(1 - \hat{p}_{CE})}{n_{1,CE}} + \frac{\hat{p}_P(1 - \hat{p}_P)}{n_{1,P}}}$$

where $\hat{p}_P$ and $\hat{p}_{CE}$ are the mean stage 1 response rates for the placebo and CE arms, respectively.

*Unconditional CIs*

For the **exact unconditional CI** we again use the $p$-value function $P(\theta)$ (based on stagewise ordering of the sample space) to find the lower and upper and bounds for the 95% CI, $\hat{\theta}_l$ and $\hat{\theta}_u$, which are the solutions to the equations $P(\hat{\theta}_l) = 0.025$ and $P(\hat{\theta}_u) = 0.975$. This admits a closed-form expression when $T = 1$, with $\hat{\theta}_l = \frac{\Phi^{-1}(\alpha/2) + Z_1}{\sqrt{I_1}}$ and $\hat{\theta}_u = \frac{\Phi^{-1}(1 - \alpha/2) + Z_1}{\sqrt{I_1}}$.

The **repeated CI (RCI)** and **adjusted asymptotic CI** are given in Section 3.1, since they are written in terms of a general stopping stage $T$.

The **parametric bootstrap CI** procedure is the same as before, except that now $\hat{p}_{CE}$ and $\hat{p}_P$ represent the stage 1 response rate estimates on the CE and placebo arm, respectively.

*Conditional CIs*

When $T = 1$, the **exact conditional CI** has the same definition as given for $CI_c$ in Section 3.1. More explicitly, for a trial stopping at stage 1, the lower and upper bounds for the CI, denoted $\hat{\theta}_l$ and $\hat{\theta}_u$, are the solutions to the equations $\alpha/2 = \frac{1 - \Phi(\sqrt{I_1}[\hat{\theta}_{obs} - \hat{\theta}_l])}{1 - \Phi(e_1 - \sqrt{I_1}\hat{\theta}_l)}$ and $1 - \alpha/2 = \frac{1 - \Phi(\sqrt{I_1}[\hat{\theta}_{obs} - \hat{\theta}_u])}{1 - \Phi(e_1 - \sqrt{I_1}\hat{\theta}_u)}$.

Similarly, the **restricted exact conditional CI** has the same definition $CI_c \cap CI_r$ with $CI_r$ as given in Section 3.1. For a trial stopping at stage 1, the lower bound of the restricted exact conditional CI is set equal to $max\{\hat{\theta}_l, (e_1 - \Phi^{-1}(1 - \alpha/2))/\sqrt{I_1}\}$, where $\hat{\theta}_u$ is the upper bound of $CI_c$.

The **conditional likelihood CI** is based on the conditional log-likelihood as given in Section 3.1. Again using the results of Fan and DeMets (2006)[22], the conditional MLE, $\hat{\theta}_c$, is the solution of the following equation when $T = 1$:

$$\hat{\theta}_c = \hat{\theta}_{obs} - \frac{\phi(e_1 - \hat{\theta}_c \sqrt{I_1})}{\sqrt{I_1}\Phi(e_1 - \hat{\theta}_c \sqrt{I_1})}$$

The conditional likelihood CI is calculated using a conditional bootstrap procedure, with full details given in Appendix A.1.

Finally, the **penalised likelihood CI** is different from the conditional likelihood CI when the trial stops at stage 1. As described in Marschner et al. (2022)[23], the penalised log-likelihood is given by $L_\lambda(\theta \,;\, z_t \,, t) = -\frac{1}{2}(z_t - \theta\sqrt{I_t})^2 - \lambda \, log \, Prob(T = t \mid \theta)$. Hence the MLE and the conditional MLE correspond to maximising the penalised log-likelihood when $\lambda = 0$ and $\lambda = 1$, respectively. For a given choice of $\lambda \in [0,1]$ this gives a penalised likelihood estimate $P(\lambda, z_t, t) = argmax_\theta \, L_\lambda(\theta \,;\, z_t \,, t)$. The choice of $\lambda$ proposed for $t = 1$ is $\lambda^* = \{\lambda \in [0,1] : P(\lambda, e_1, 1) = 0\}$. With this choice, the penalised MLE is defined as $\hat{\theta}_p = P(\lambda^*, z_1, 1)$. The penalised likelihood CI is then calculated using the same bootstrap procedure as above for the conditional likelihood CI, with the bootstrap conditional MLE $\hat{\theta}_c^{(b)}$ replaced by the penalised MLE $\hat{\theta}_p^{(b)}$. Note that by conditioning the bootstrap sampling on early stopping, each bootstrap replication satisfies $\hat{\theta}_p^{(b)} > 0$, which therefore guarantees that the associated CI lies above zero, consistent with the decision to stop the study early for benefit.

### 3.3 Results from the MUSEC trial

Table 2 gives the values of all of the CIs described in Section 3.1, calculated using the observed data and OBF stopping boundaries from the MUSEC trial, with Figure 1 giving the graphical representation. When there is an associated point estimate (as described above in Section 2.1), this is also shown. For the methods requiring repeated sampling/simulation (i.e., the unconditional parametric bootstrap CI, unconditional randomisation-based CI and conditional (penalised) likelihood CI), we used $N = 10^6$ trial replicates.

| Type of CI | CI method | Point estimate | 95% two-sided CI | CI width |
|---|---|---|---|---|
| **Standard/naive** | Wald test | 0.137 [overall MLE] | (0.040, 0.234) | 0.194 |
| **Unconditional** | Exact | 0.134 [MUE] | (0.034, 0.234) | 0.200 |
| | Repeated | - | (0.037, 0.237) | 0.199 |
| | Adjusted asymptotic | 0.137 | (0.039, 0.235) | 0.196 |
| | Parametric bootstrap | 0.143 | (0.041, 0.253) | 0.212 |
| | Randomisation-based | 0.130 | (0.033, 0.226) | 0.194 |
| **Conditional** | Exact | 0.185 [conditional MUE] | (0.052, 0.358) | 0.306 |
| | Restricted exact | 0.185 [conditional MUE] | (0.052, 0.269) | 0.217 |
| | (Penalised) likelihood | 0.191 [conditional MLE] | (0.034, 0.304) | 0.271 |

*Table 2: Confidence intervals (and associated point estimates) calculated using the observed data and O'Brien-Fleming efficacy stopping boundaries from the MUSEC trial.*

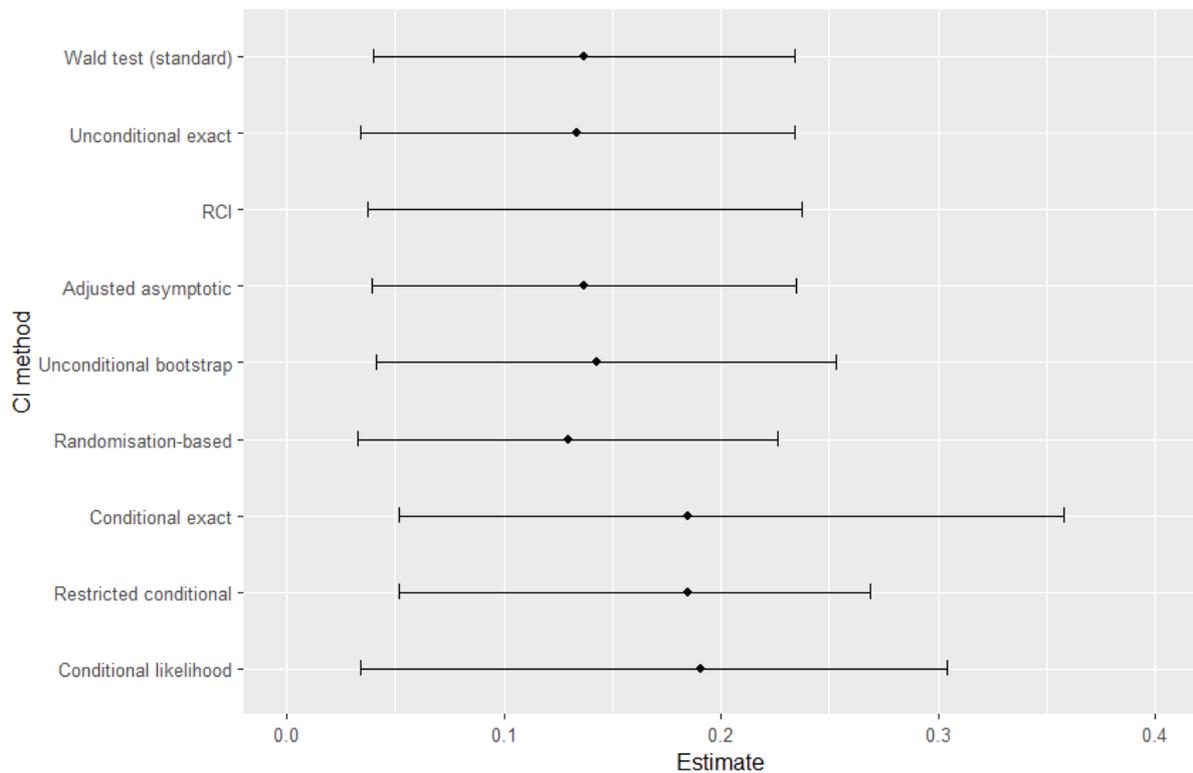

Figure 1: *Graphical representation of confidence intervals (and associated point estimates) calculated using the observed data from the MUSEC trial. RCI = Repeated Confidence Interval.*

The Wald (standard) CI is (0.040, 0.234) with a width of 0.194, and is the comparator for all the other CIs in Table 2, since it is the conventional end-of-trial CI. Starting with the unconditional CIs, all of them are very similar to the standard CI, with the exception of the simple bootstrap method which is wider (an increase of 9%) than the standard CI. In contrast, the conditional CIs all are substantially wider than the standard CI, with an increase of 58%, 12% and 40% for the conditional exact, restricted exact and (penalised) likelihood CIs, respectively. This reflects the loss of information associated with conditioning on the stopping stage $T = 2$. The conditional point estimators are also substantially higher than the unconditional point estimators. This upward correction is intuitive from a conditional perspective: there is downward 'selection pressure' on the MLE calculated at the end of stage 1, since if this is large then the trial does not continue to stage 2. For the conditional CIs, this is also reflected in their upper confidence limits being substantially higher compared with the standard CI. Finally, there is a marked asymmetry in the restricted exact CI around its associated point estimate, reflecting how the upper confidence limit of the conditional exact CI is adjusted.

For the MUSEC trial data, the use of different methods can give noticeably different CIs for the treatment effect, particularly when considering a conditional versus unconditional perspective. This could influence the interpretation of the trial results, and highlights the importance of pre-specifying which CI(s) will be reported following an AD. The choice of CI(s) will depend on what the researchers wish to achieve regarding the estimand[24] in question.

There will be pros and cons for the different CI methods, which we explore further in the simulation study in Section 4. We also note that there is a strong link between design and estimation - the CIs above depend on the design of the trial, and would be different if (for example) the design had also included futility stopping boundaries.

## 4. Simulation study

### 4.1 Simulation set-up

Since the CIs calculated above are one realisation of the trial data given the trial design, in this Section we carry out a simulation study to investigate the performance of the CIs under different scenarios. Note that we have *not* used assumed values for the underlying treatment effects to calculate the CIs given in Table 2. As can be seen from the formulae in Section 3.1, these CIs only depend on the observed data and efficacy stopping boundary.

To demonstrate the properties of the CIs when averaged over many trial realisations following the two-stage design of the MUSEC trial, we ran simulations under different values of $p_{CE}$, the assumed true value of the response rate for the CE arm. For simplicity, we kept the value of $p_P$, the assumed true value of the response rate for the placebo arm, equal to the value observed in the MUSEC trial, i.e. $p_P = \widehat{p_p} = 21/134 \approx 0.157$.

We used the following procedure for our trial simulations:

1) Given the assumed value for $p_{CE}$, denoted $p_{CE}^*$, generate $N$ stage 1 trial replicates $S_{1,CE}^{(1)}, \ldots, S_{1,CE}^{(N)}$ and $S_{1,P}^{(1)}, \ldots, S_{1,P}^{(N)}$, where $S_{1,CE}^{(i)} \sim Bin(n_{1,CE}, p_{CE}^*)$ and $S_{1,P}^{(i)} \sim Bin(n_{1,P}, \hat{p}_P)$.

2) For $i = 1, \ldots, N$ calculate the bootstrap standardised stage 1 test statistic $Z_1^{(i)}$ from the bootstrap values $S_{1,CE}^{(i)}$ and $S_{1,P}^{(i)}$.
    a) If $Z_1^{(i)} > e_1$ then calculate all CIs (conditional and unconditional) with $T = 1$.
    b) Otherwise, generate a stage 2 trial replicate $S_{2,CE}^{(i)}$ and $S_{2,P}^{(i)}$, where $S_{2,CE}^{(i)} \sim Bin(n_{CE} - n_{1,CE}, p_{CE}^*)$ and $S_{2,P}^{(i)} \sim Bin(n_p - n_{1,P}, \hat{p}_P)$. Then calculate all CIs (conditional and unconditional) with $T = 2$.

For each value of $p_{CE}$, we simulated $N = 10^5$ trial replicates. For the CI methods requiring a bootstrap procedure, we used $B = 10^4$ bootstrap samples. For the randomisation-based CI, even with $10^4$ samples we still frequently ran into the issue of $p_{adjusted} = 0$. Hence we did not consider this CI method further in the simulation study.

For each of the CI methods considered, we evaluated the following properties:

1) Mean coverage
2) Mean width and standard error (se)
3) Consistency
4) Lower coverage: probability that the lower confidence limit is above the true value of $\theta$, denoted $P(L(X) > \theta)$.
5) Upper coverage: probability that the upper confidence limit is below the true value of $\theta$, denoted $P(U(X) < \theta).$

6)

## 4.2 Simulation results

Table 3 shows the overall (unconditional) simulation results with the true success rates $(p_p, p_{CE})$ on the two arms equal to the overall observed means in the MUSEC trial i.e., $p_p = 21/134 \approx 0.157$ and $p_{CE} = 42/143 \approx 0.294$. The probability of stopping early for efficacy in stage 1 is 0.308. Note that with $N = 10^5$ trial replicates, the Monte Carlo standard error for the coverage and consistency results is less than 0.0016.

**Overall (unconditional) results**

| Type of CI | CI method | Coverage | Mean width (se) | Consistency | $P(L(X) > \theta)$ | $P(U(X) < \theta)$ |
|---|---|---|---|---|---|---|
| **Standard/naive** | Wald test | 0.945 | 0.203 (0.017) | 0.989 | 0.029 | 0.026 |
| **Unconditional** | Exact | 0.952 | 0.211 (0.019) | 0.998 | 0.022 | 0.026 |
|  | Repeated | 0.973 | 0.240 (0.063) | 1.000 | 0.002 | 0.025 |
|  | Adjusted asymptotic | 0.945 | 0.205 (0.020) | 0.985 | 0.022 | 0.034 |
|  | Parametric bootstrap | 0.926 | 0.210 (0.010) | 0.988 | 0.056 | 0.018 |
| **Conditional** | Exact | 0.954 | 0.386 (0.268) | 0.732 | 0.024 | 0.022 |

| | | | | | |
|---|---|---|---|---|---|
| Restricted exact | 0.954 | 0.227 (0.041) | 0.985 | 0.024 | 0.022 |
| Likelihood | 0.988 | 0.485 (0.363) | 0.693 | 0.002 | 0.010 |
| Penalised likelihood | 0.988 | 0.271 (0.028) | 0.992 | 0.002 | 0.010 |

*Table 3: Simulation results showing the performance of various CIs with $p_p = 21/134 \approx 0.157$ and $p_{CE} = 42/143 \approx 0.294$. There were $10^5$ trial replicates. The probability of stopping at stage 1 is 0.308.*

Starting with the coverage of the CIs, the standard CI has a very slight undercoverage, which is caused by two factors: 1) the distributional assumptions underlying the Wald CI are no longer met due to the stopping rule, and 2) the quality of the (asymptotic) normal approximation used for binomial outcomes. As expected, the exact unconditional CI attains the nominal coverage of 95% (within Monte Carlo error). In contrast, the RCI has conservative coverage, which is driven by the trial replicates that stop early at stage 1 (see the following tables). The adjusted asymptotic CI has the same coverage as the standard CI. The parametric bootstrap CI has particularly low coverage of <93% in this scenario.

Turning to the conditional estimators, both the conditional exact CI and conditional restricted exact CI attain the nominal coverage of 95% (with a very slight overcoverage). This is to be expected, since if a CI attains (or exceeds) the nominal coverage conditional on stopping at stage 1 and conditional on continuing to stage 2, then it will also attain (or exceed) the nominal coverage unconditionally (averaged over the two stopping possibilities). In contrast, the conditional likelihood and penalised likelihood have a notably conservative coverage, which is at least partly due to the choice of the conditional bootstrap procedure (using the bias-corrected bootstrap, for example, would give different results but is out of scope of this paper).

Looking at the mean CI width, the CI methods with higher mean coverage compared with the standard CI also have a higher mean width. However, even though the parametric bootstrap CI has a lower coverage than the standard CI, ist also has a slightly higher mean width. The mean widths of the unconditional exact, adjusted asymptotic and parametric bootstrap CIs are all within +4% of the mean width of the standard CI. In contrast, the RCI has a substantial increase of +17%. As for the conditional CIs, what is striking is the huge increases in mean width for the conditional exact and conditional likelihood CIs, of +90% and +139%, respectively. These are also accompanied with very high standard errors, reflecting a very high variability in the confidence limits. Such results have previously been noted in the literature[22,23,25]. In contrast, the conditional restricted exact and penalised likelihood CIs have mean widths much closer to that of the standard CI, although there is still an increase of +12% and +33%, respectively. Again, this reflects the loss of information associated with conditioning on the stopping stage.

In terms of consistency, the standard CI has consistency just below 99%, with instances of non-consistent CIs being caused by the same reasons as for the undercoverage. The unconditional exact CI has a coverage just below 100%, with the non-consistency caused by the the quality of the (asymptotic) normal approximation used for binomial outcomes, see also Lloyd (2021). The RCI is the only CI method with 100% consistency. Both the adjusted asymptotic and unconditional parametric bootstrap CIs have very similar consistency to the standard CI. For the conditional CIs, again it is striking how low the consistency is for the conditional exact and conditional likelihood CIs, which is driven by how they are so wide (on average) that they often will include zero even when the null hypothesis of no treatment effect is rejected. In contrast, the conditional restricted exact and penalised likelihood CIs have consistencies much closer to 100%, comparable to the consistency of the standard CI.

Finally, looking at the upper and lower coverages, these are approximately equal for the standard, unconditional exact, conditional exact and conditional restricted exact CIs. The upper coverage is (sometimes substantially) greater than the lower coverage for the RCI, adjusted asymptotic, conditional likelihood and penalised likelihood CIs. The unconditional parametric bootstrap CI is the only method to have the lower coverage higher than the upper coverage.

In order to show more clearly the differences between the CIs in repeated realisations of the MUSEC trial, Figure 2 shows the CIs from the first 10 simulation replicates of the simulation study. Note that in simulation replicates 2, 3, 6, 8 and 10, the trial stopped early at stage 1 (and hence with rejection of the null hypothesis). In simulation replicates 1, 4, 5, 7 and 9, the trial continued to stage 2, with rejection of the null hypothesis in simulation replicates 1, 7 and 9.

Looking first at coverage (i.e., whether the CI contains the true value of the treatment difference, shown by the red horizontal line), in simulation replicates 1, 2, 3, 6, 7, 8 and 9, all CI methods contain the true value, whereas in simulation replicate 5 all CI methods do not contain the true value. However, in simulation replicate 4 we can see a discrepancy in the coverage, with the standard, unconditional exact, RCI and adjusted asymptotic CIs not containing the true value, whereas the other CIs do contain the true value. In simulation replicate 10, all CI methods contain the true value except for the bootstrap CI. Similarly, in terms of consistency, in simulation replicates 2, 3, 6 and 8, the conditional exact and conditional likelihood CIs contain zero (with the conditional likelihood CI additionally containing zero in simulation replicate 10), despite the null hypothesis of no treatment effect being rejected.

Finally, in terms of CI width, in all of the simulation replicates where the trial stopped early in stage 1 (apart from simulation replicate 10), it is striking how much wider both the conditional exact and conditional likelihood CIs are compared to any of the others. In the trial replicates that continue to stage 2, the CI widths are much more similar, although the conditional CIs are wider than the unconditional ones (as would be expected).

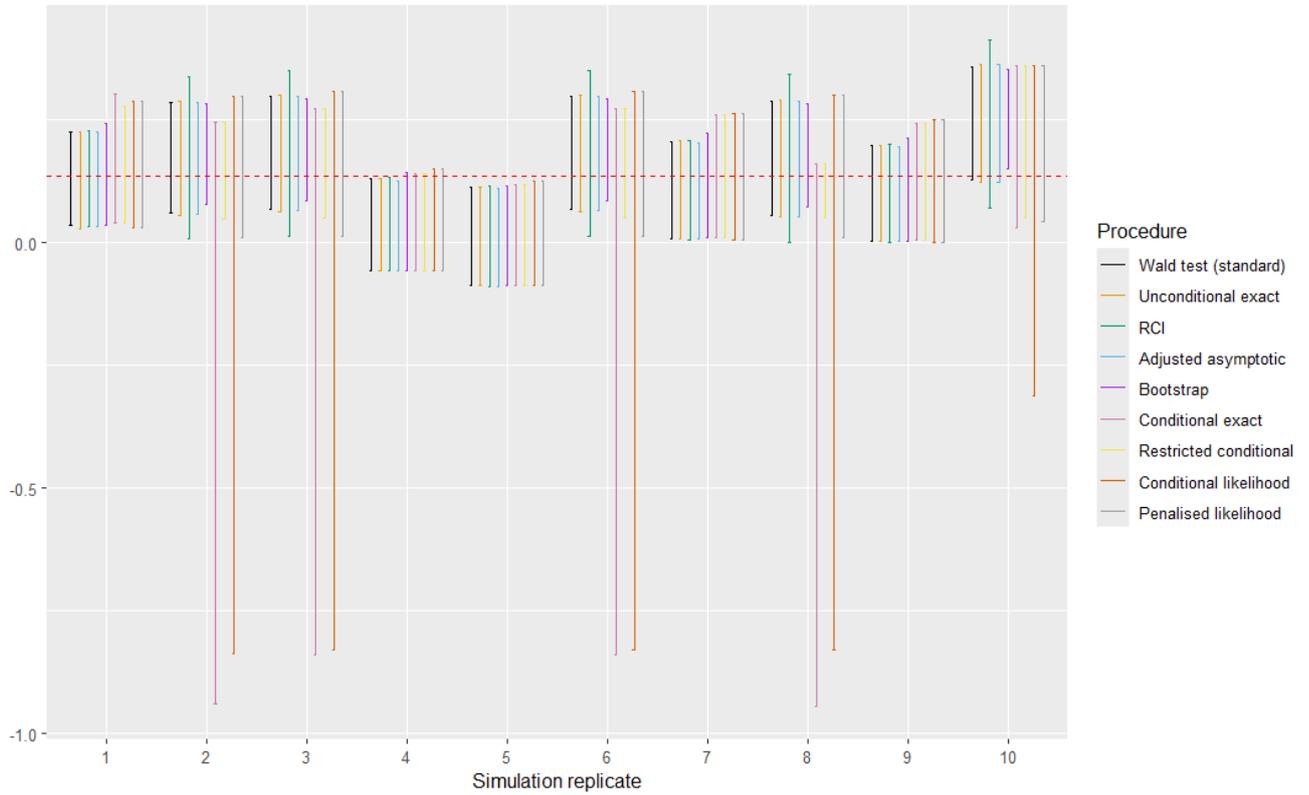

Figure 2: *Confidence intervals from the first 10 simulation replicates of the simulation study. The horizontal red dashed line shows the true value of the treatment difference. RCI = Repeated Confidence Interval.*

**Results conditional on stopping at stage 1**

Apart from the overall (unconditional) results, it is informative to also report results conditional on the stopping stage of the trial. Table 4 shows the simulation results conditional on stopping early at stage 1, with the true success rates ($p_p$, $p_{CE}$) on the two arms again equal to the overall observed means in the MUSEC trial. With a probability of early stopping of 0.308, these results are based on $3.08 \times 10^4$ simulation replicates, giving a Monte Carlo standard error of less than 0.0028 for the coverage and consistency results.

| Type of CI | CI method | Coverage | Mean width (se) | Consistency | $P(L(X) > \theta)$ | $P(U(X) < \theta)$ |
|---|---|---|---|---|---|---|
| **Standard/naive** | Wald test | 0.907 | 0.225 (0.010) | 1.000 | 0.093 | 0.000 |
| **Unconditional** | Exact | 0.930 | 0.234 (0.011) | 1.000 | 0.070 | 0.000 |
|  | Repeated | 0.995 | 0.334 (0.015) | 1.000 | 0.005 | 0.000 |
|  | Adjusted asymptotic | 0.930 | 0.232 (0.011) | 1.000 | 0.070 | 0.000 |
|  | Parametric bootstrap | 0.820 | 0.207 (0.010) | 1.000 | 0.180 | 0.000 |
| **Conditional** | Exact | 0.970 | 0.673 (0.331) | 0.179 | 0.017 | 0.013 |
|  | Restricted exact | 0.970 | 0.242 (0.054) | 1.000 | 0.017 | 0.013 |
|  | Likelihood | 0.995 | 0.992 (0.229) | 0.030 | 0.005 | 0.000 |
|  | Penalised likelihood | 0.993 | 0.298 (0.017) | 1.000 | 0.007 | 0.000 |

*Table 4: Simulation results showing the performance of various CIs with $p_p = 21/134 \approx 0.157$ and $p_{CE} = 42/143 \approx 0.294$, conditional on the trial stopping early at stage 1. There were $10^5$ trial replicates. The probability of stopping at stage 1 is 0.308.*

Conditional on stopping at stage 1, the coverage of the standard CI is substantially below the nominal, at less than 91%. The coverage of the unconditional exact and adjusted asymptotic CIs also decreases to 93%, while the parametric bootstrap has the largest drop to only 82%. In contrast, the RCI has a very conservative coverage of 99.5%. These results demonstrate that even if unconditionally the unconditional CIs may have near nominal coverage, the conditional coverage properties can be poor. In contrast, all of the conditional CI methods (as expected) have coverage above the nominal 95%, with the conditional exact and restricted exact CIs having conservative coverage of 97%, and the conditional likelihood and penalised likelihood CIs having a very conservative coverage comparable to the RCI.

In terms of mean CI width, similar patterns are seen as for the unconditional results. Some noticeable features are that the RCI has a substantially higher mean width than either the conditional restricted exact or penalised likelihood CIs. The very large mean widths of the conditional exact and conditional likelihood CIs are even more extreme conditional on stopping at stage 1, with increases of +199% and +340%, respectively, compared with the mean width of the standard CI. This agrees with the literature[22,23] that does not recommend the use of these two methods when a group sequential trial stops early for benefit. The very large mean widths also correspond with very low consistencies of the test decision (which is always to reject the null). In contrast, all other CI methods have 100% consistency (with this holding by design for the RCI and penalised likelihood CI). Finally, all CI methods except for the conditional exact and restricted exact CIs have an upper coverage of zero.

**Conditional on continuing to stage 2**

Table 5 shows the simulation results conditional on continuing to stage 2, with the true success rates ($p_p$, $p_{CE}$) on the two arms again equal to the overall observed means in the MUSEC trial. With a probability of continuing to stage 2 of 69.2%, these results are based on $6.92 \times 10^4$ simulation replicates, giving a Monte Carlo standard error of less than 0.0019 for the coverage and consistency results.

| Type of CI | CI method | Coverage | Mean width (se) | Consistency | $P(L(X) > \theta)$ | $P(U(X) < \theta)$ |
|---|---|---|---|---|---|---|
| **Standard/naive** | Wald test | 0.962 | 0.193 (0.008) | 0.984 | 0.000 | 0.038 |
| **Unconditional** | Exact | 0.962 | 0.200 (0.010) | 0.998 | 0.001 | 0.038 |
| | Repeated | 0.963 | 0.198 (0.008) | 1.000 | 0.000 | 0.036 |
| | Adjusted asymptotic | 0.951 | 0.193 (0.008) | 0.978 | 0.000 | 0.049 |
| | Parametric bootstrap | 0.973 | 0.211 (0.009) | 0.983 | 0.000 | 0.027 |
| **Conditional** | Exact | 0.947 | 0.258 (0.040) | 0.978 | 0.027 | 0.025 |
| | Restricted exact | 0.947 | 0.220 (0.031) | 0.978 | 0.027 | 0.025 |
| | (Penalised) likelihood | 0.985 | 0.258 (0.023) | 0.989 | 0.000 | 0.015 |

*Table 5: Simulation results showing the performance of various CIs with $p_p = 21/134 \approx 0.157$ and $p_{CE} = 42/143 \approx 0.294$, conditional on the trial continuing to stage 2. There were $10^5$ trial replicates. The probability of continuing to stage 2 is 0.692.*

Conditional on continuing to stage 2, the standard CI and all the unconditional CIs now have slightly conservative coverage of around 96 - 97% (with the exception of the adjusted asymptotic CI which achieves the nominal 95% coverage). In contrast, the coverage of conditional exact and restricted exact CIs is just below the nominal 95%. The conditional likelihood CI (which is the same as the penalised likelihood CI in this case) has the most conservative coverage of 98.5%. What is noticeable is that the mean widths of the conditional exact and conditional (penalised) likelihood CIs are much lower than the mean widths conditional on early stopping at stage 1, with increases of +14% and +34% compared with the standard CI.

In terms of consistency, there is a drop of up to around 2% (compared with the 100% consistency conditional on stopping at stage 1) for the standard CI, conditional restricted exact and all unconditional CIs, with the exception of the RCI which maintains 100% consistency and the unconditional exact CI which has almost 100% consistency (with inconsistency again caused by the normal approximation). Meanwhile, the consistency of the conditional exact and conditional likelihood CIs are around 98 - 99%, compared with only 18% and 3% consistency, respectively, conditional on stopping at stage 1. Finally, all CI methods except for the conditional exact and restricted exact CIs now have a *lower* coverage of zero.

If however we run the simulations with a higher true success rates for $p_{CE}$ i.e., $p_{CE} = 42/143 + 0.08 \approx 0.374$ so that the probability of continuing to stage 2 is only 0.239, the coverage results conditional on continuing to stage 2 look rather different. Table 6 shows these simulation results, with the Monte Carlo standard error for the coverage and consistency results less than 0.0032.

| Type of CI | CI method | Coverage | Mean width (se) | Consistency | $P(L(X) > \theta)$ | $P(U(X) < \theta)$ |
|---|---|---|---|---|---|---|
| **Standard/naive** | Wald test | 0.897 | 0.202 (0.007) | 0.994 | 0.000 | 0.103 |
| **Unconditional** | Exact | 0.902 | 0.218 (0.012) | 0.997 | 0.000 | 0.098 |
|  | Repeated | 0.910 | 0.209 (0.007) | 1.000 | 0.000 | 0.090 |
|  | Adjusted asymptotic | 0.895 | 0.205 (0.007) | 0.992 | 0.000 | 0.105 |
|  | Parametric bootstrap | 0.959 | 0.221 (0.008) | 0.995 | 0.000 | 0.041 |
| **Conditional** | Exact | 0.945 | 0.310 (0.043) | 0.989 | 0.031 | 0.024 |
|  | Restricted exact | 0.945 | 0.203 (0.055) | 0.989 | 0.031 | 0.024 |
|  | (Penalised) likelihood | 0.987 | 0.288 (0.019) | 0.996 | 0.000 | 0.013 |

*Table 6: Simulation results showing the performance of various CIs with $p_p = 21/134 \approx 0.157$ and $p_{CE} = 42/143 + 0.08 \approx 0.374$, conditional on continuing to stage 2. There were $10^5$ trial replicates. The probability of continuing to stage 2 is 0.239.*

This time, the standard CI and all the unconditional CIs, including the RCI, now have coverage substantially less than the nominal 95%, ranging from 89 - 91%. The exception is the parametric bootstrap CI, which has a slightly conservative coverage (96%). In contrast, the coverage of conditional exact and restricted exact CIs remains just below the nominal 95%, while the coverage of the conditional (penalised) likelihood CI remains rather conservative (almost 99%). Again these results demonstrate that unconditional CIs can have poor coverage properties conditional on the stopping stage (regardless of whether the trial is stopped early or continues). Tables showing the simulation results unconditionally and conditional on early stopping at stage 1 for this choice of values for $p_p$ and $p_{CE}$ can be found in Appendix A.2.

**Simulation results across a range of values for $p_{CE}$**

The results given in Tables 4-6 above are only for a particular fixed value of $p_{CE}$ and so we now explore the performance of the various CI methods across a range of values of $p_{CE}$ from $\hat{p}_{CE} - 0.07 \approx 0.224$ to $\hat{p}_{CE} + 0.14 \approx 0.434$, which corresponds to a probability of early stopping ranging from 0.05 to 0.94, as seen in Figure 3 below.

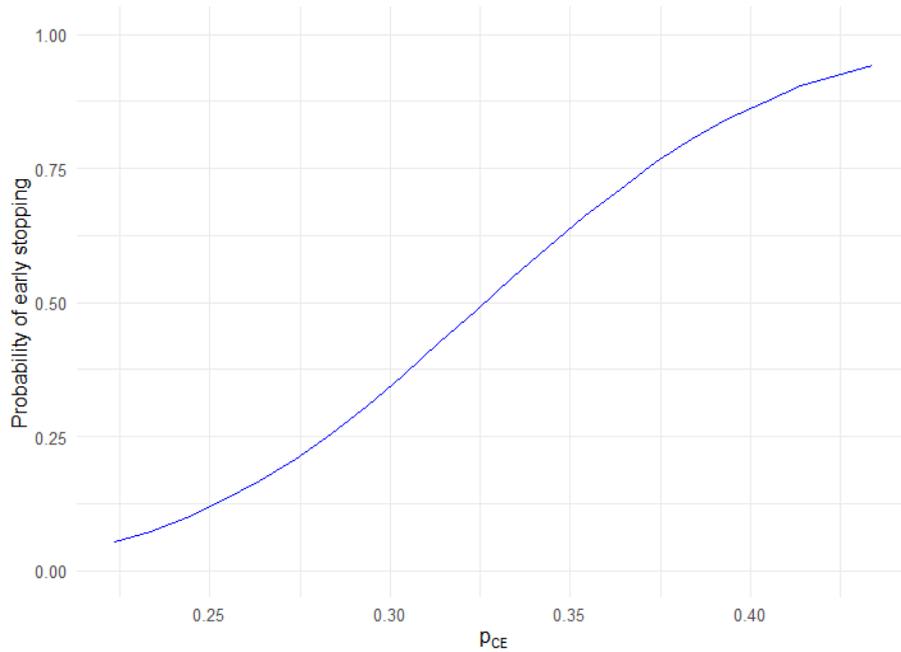

Figure 3: *Probability of early stopping at stage 1 as the value of $p_{CE}$ varies from $\hat{p}_{CE} - 0.07 \approx 0.224$ to $\hat{p}_{CE} + 0.14 \approx 0.434$. The value of $p_p = 21/134 \approx 0.157$.*

For each value of $p_{CE}$ (22 in total), we ran $N = 10^5$ trial replicates for each of the CI methods, except for the standard CI where we ran $N = 10^6$ trial replicates in order to more accurately assess whether there was any undercoverage. For the CI methods requiring a bootstrap procedure (unconditional parametric bootstrap, conditional likelihood and conditional penalised likelihood), we again used $B = 10^4$ bootstrap samples. Figure 4 shows the overall (unconditional) coverage of the CI methods as a function of $p_{CE}$. The shaded areas around the lines for the CI methods correspond to $\pm 1.96$ times the Monte Carlo standard error. The red dashed line denotes the nominal 95% coverage.

Starting with the standard CI, the coverage is (just) below the nominal 95% for most of the range of $p_{CE}$ except for $p_{CE} > 0.40$ when it goes just above 95%. In contrast, the unconditional exact CI has coverage (just) above the nominal 95% for the whole range of $p_{CE}$, with the coverage becoming increasingly conservative (up to 96%) as $p_{CE}$ increases. The RCI has rather conservative coverage for the whole range of $p_{CE}$, with the coverage also becoming increasingly conservative (>98%) as $p_{CE}$ increases. The adjusted asymptotic CI has very similar undercoverage to the standard CI for $p_{CE} < 0.33$ but then has increasingly conservative coverage as $p_{CE}$ increases above 0.33. The unconditional parametric bootstrap CI has rather low coverage (<93%) for $p_{CE} < 0.32$ but quickly switches to having increasingly conservative coverage for $p_{CE} > 0.34$. These results demonstrate that whether a particular CI method has the correct coverage can strongly depend on the true (unknown) underlying parameter values. The conditional CIs all have conservative coverage, but the conditional exact and conditional restricted exact CIs have coverage close to the nominal (95-96%). In contrast, the conditional likelihood and penalised likelihood CI have rather conservative coverage (98-99%).

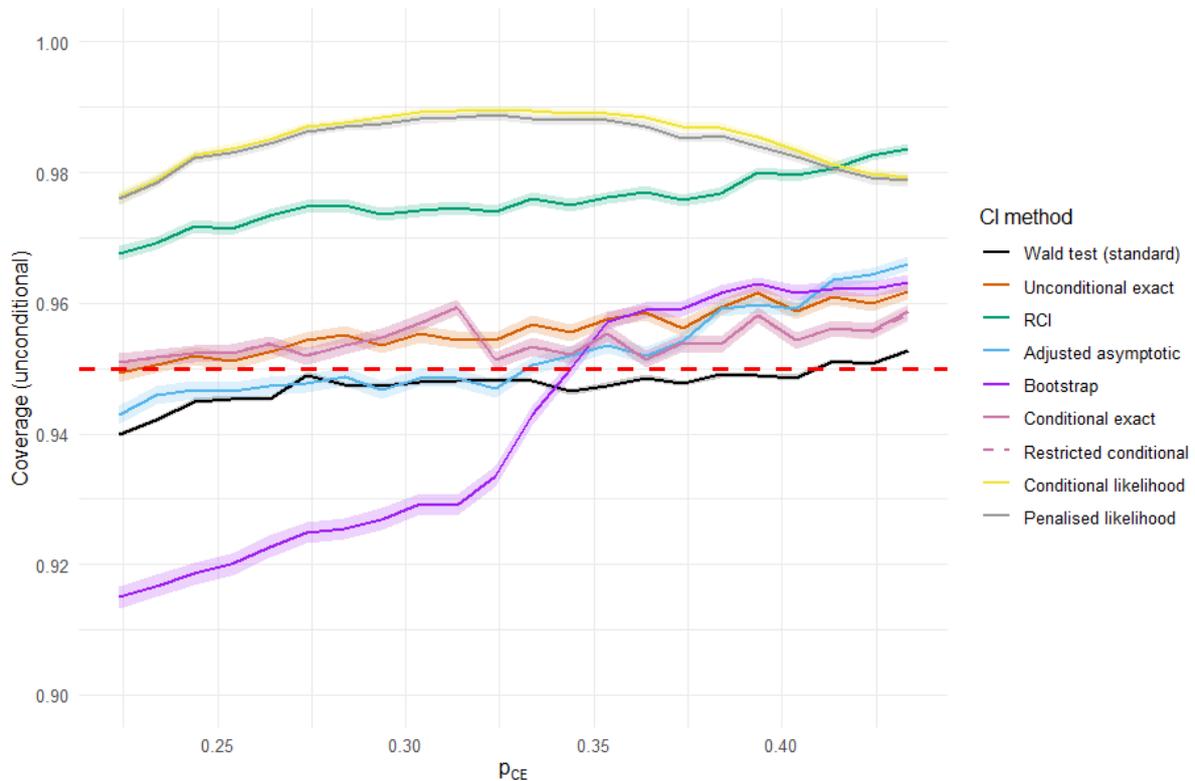

Figure 4: *Coverage as the value of $p_{CE}$ varies from $\hat{p}_{CE} - 0.07 \approx 0.224$ to $\hat{p}_{CE} + 0.14 \approx 0.434$. The value of $p_p = 21/134 \approx 0.157$. The shaded areas around the lines for the CI methods correspond to $\pm 1.96$ times the Monte Carlo standard error. The red dashed line denotes the nominal 95% coverage. The restricted conditional and conditional exact lines are almost completely overlapping.*

Figure 5 shows the (unconditional) mean width of the CI methods as a function of $p_{CE}$. The unconditional exact, adjusted asymptotic and unconditional parametric bootstrap CIs all have similar mean width as the standard CI (within $\pm 6\%$). The conditional restricted exact CI has a slightly higher mean width than the standard CI (between 8-13%), with the penalised likelihood CI having a mean width between 23-33% higher than the standard CI. In contrast, the conditional exact and conditional likelihood CI mean widths are dramatically larger than the standard CI, up to 106% and 203% larger, respectively. Finally, the RCI has a similar mean width as the standard CI for small values of $p_{CE}$ but becomes substantially higher as $p_{CE}$ increases (up to 43% larger).

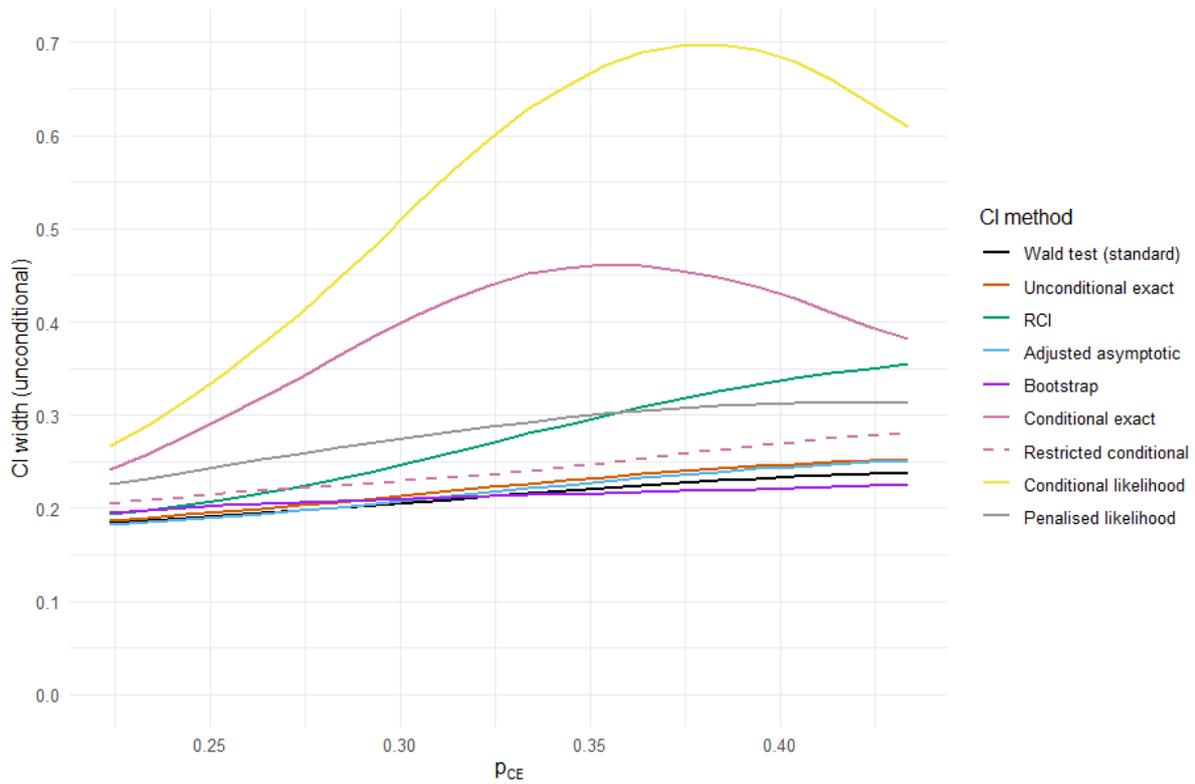

Figure 5: *CI width as the value of $p_{CE}$ varies from $\hat{p}_{CE} - 0.07 \approx 0.224$ to $\hat{p}_{CE} + 0.14 \approx 0.434$. The value of $p_p = 21/134 \approx 0.157$.*

As before, it is informative to also report results conditional on the stopping stage of the trial. Figure 6 shows the coverage of the CI methods conditional on early stopping at stage 1 as a function of $p_{CE}$. This time, for smaller values of $p_{CE}$ the coverage of the standard CI, adjusted asymptotic CI, unconditional exact and unconditional bootstrap CI is very low. The standard CI coverage can even be less than 50%. Note though that the lowest coverages are also achieved when there are the lowest probabilities of actually stopping at stage 1. The coverage of the standard CI does go above the nominal 95% when $p_{CE} > 0.33$. In contrast, all the conditional CIs and the RCI have conservative coverage throughout the range of $p_{CE}$, with the conditional exact and restricted exact CIs being (much) closer to the nominal level of coverage compared to the RCI, conditional likelihood and penalised likelihood CI.

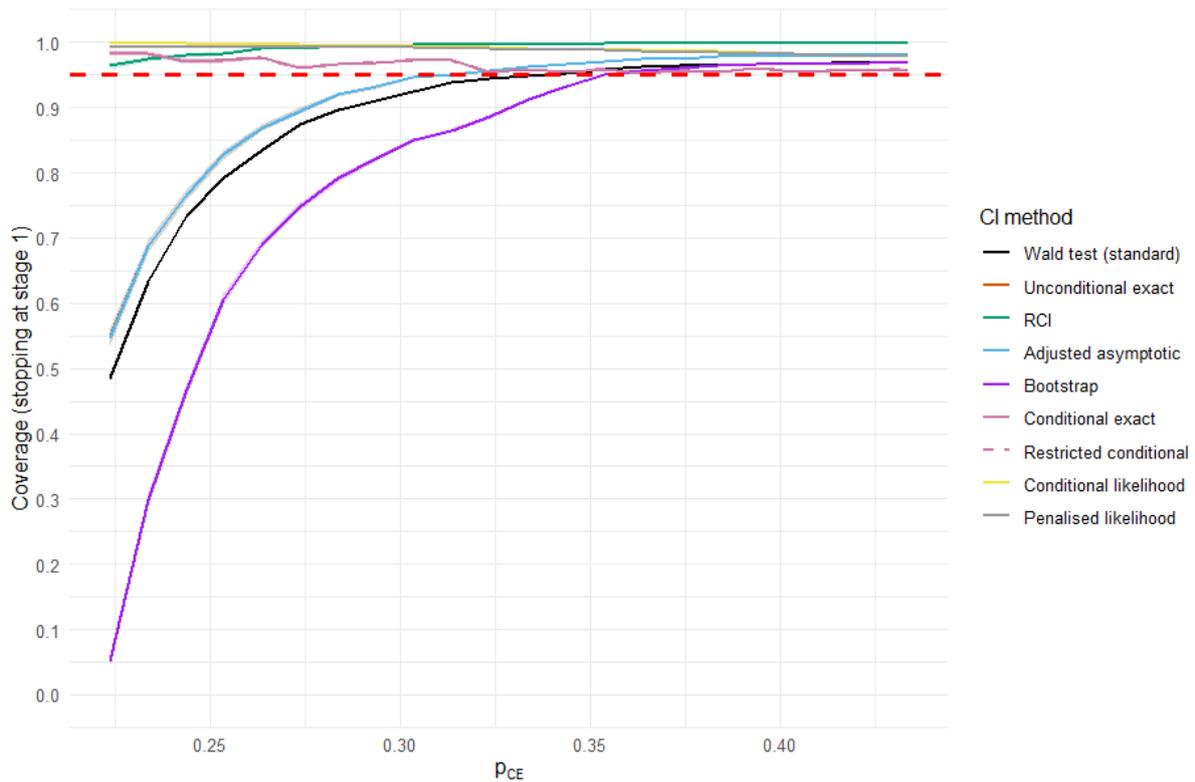

Figure 6: *Coverage conditional on early stopping at stage 1 as the value of $p_{CE}$ varies from $\hat{p}_{CE} - 0.07 \approx 0.224$ to $\hat{p}_{CE} + 0.14 \approx 0.434$. The value of $p_p = 21/134 \approx 0.157$. The shaded areas around the lines for the CI methods correspond to $\pm 1.96$ times the Monte Carlo standard error. The red dashed line denotes the nominal 95% coverage. The unconditional exact and adjusted asymptotic lines completely overlap, as do the restricted conditional and conditional exact lines. The conditional likelihood and penalised likelihood lines are also very close together.*

Figure 7 shows the mean width of the CI methods conditional on early stopping at stage 1 as a function of $p_{CE}$. There are similar patterns as for the unconditional results, except that the RCI now has substantially greater mean width than the standard CI across the range of $p_{CE}$ values (consistently between 47-52% larger). Also, the extreme widths for the conditional exact and conditional likelihood CIs are even more striking for smaller values of $p_{CE}$. This reflects the results in the literature that these CIs have poor properties conditional on early stopping, with much better properties seen with the conditional restricted exact and penalised likelihood CIs.

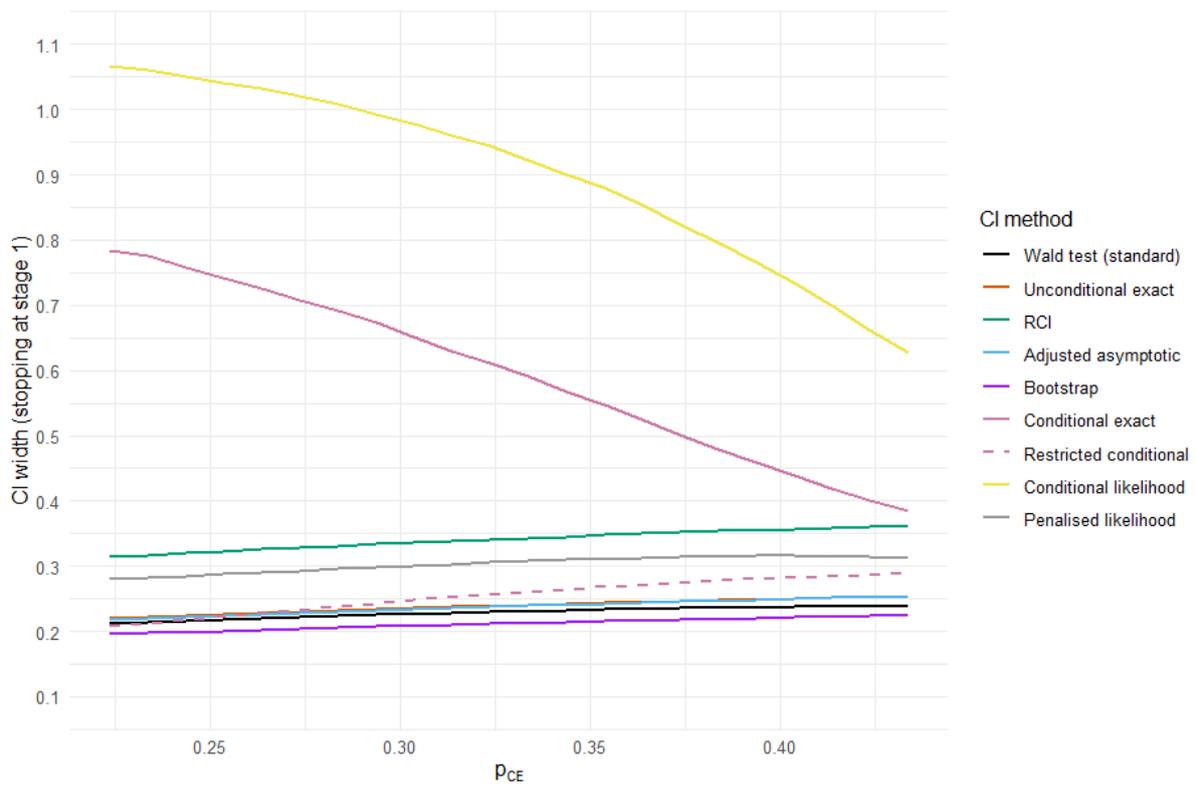

Figure 7: *CI width conditional on early stopping at stage 1 as the value of $p_{CE}$ varies from $\hat{p}_{CE} - 0.07 \approx 0.224$ to $\hat{p}_{CE} + 0.14 \approx 0.434$. The value of $p_p = 21/134 \approx 0.157$. The unconditional exact and adjusted asymptotic lines are almost completely overlapping.*

Figure 8 shows the coverage of the CI methods conditional on continuing to stage 2 as a function of $p_{CE}$. For larger values of $p_{CE}$ the coverage of the standard, adjusted asymptotic, unconditional exact and unconditional bootstrap CI is very low. The coverage of the standard CI can be below 70%. This time, the RCI is no longer conservative throughout the whole range of larger values of $p_{CE}$, with also a very low coverage for larger values of $p_{CE}$ (as low as 73%). These lowest coverages are achieved when there are the lowest probabilities of actually continuing to stage 2. The coverage of the standard CI is above the nominal 95% when $p_{CE} < 0.33$. The conditional likelihood CI (which is the same as the penalised likelihood CI) and the RCI again have conservative coverage throughout the range of $p_{CE}$. This time the conditional exact and restricted exact CIs essentially match the nominal 95% level of coverage.

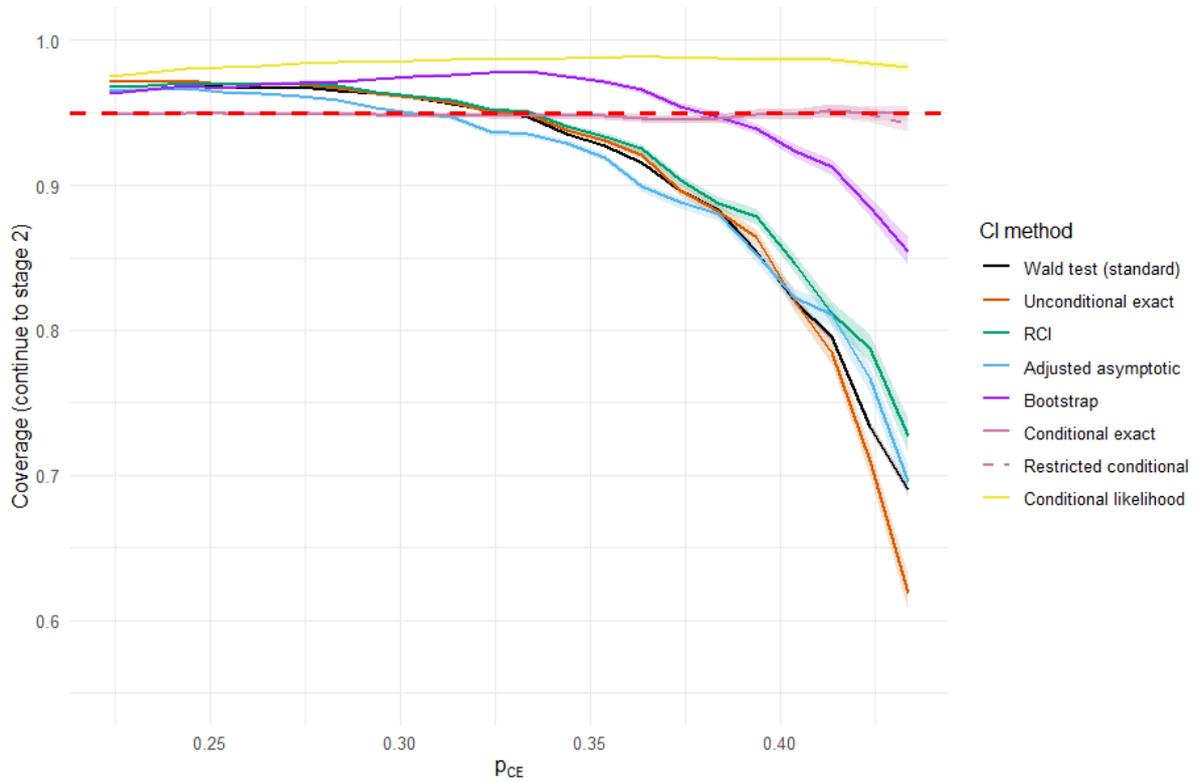

Figure 8: *Coverage conditional on continuing to stage 2 as the value of $p_{CE}$ varies from $\hat{p}_{CE} - 0.07 \approx 0.224$ to $\hat{p}_{CE} + 0.14 \approx 0.434$. The value of $p_p = 21/134 \approx 0.157$. The shaded areas around the lines for the CI methods correspond to $\pm 1.96$ times the Monte Carlo standard error. The red dashed line denotes the nominal 95% coverage. The conditional exact and restricted conditional lines are almost completely overlapping.*

Finally, Figure 9 shows the mean width of the CI methods conditional on continuing to stage 2 as a function of $p_{CE}$. The adjusted asymptotic CI has a very similar mean width as the standard CI (within $\pm 2\%$). This time though, the RCI also has a similar mean width as the standard CI, only up to 4% greater. Meanwhile, the unconditional exact CI has an increasingly large mean width compared with the standard CI as $p_{CE}$ increases (up to 12% larger). The unconditional bootstrap CI has a consistently higher mean width than the standard CI of between 7-9%. The mean width of the conditional restricted exact CI has an interesting pattern, being up to 15% greater than the standard CI for $p_{CE} < 0.38$, but then having a smaller mean width than the standard CI for $p_{CE} > 0.38$, with even up to a 16% decrease. In contrast, the conditional exact and conditional likelihood CI mean widths remain much larger than the standard CI, up to 64% and 46% larger, respectively. This implies that the penalised likelihood CI also has much larger widths than the standard CI conditional on continuing to stage 2 (as it is equal to the conditional likelihood CI).

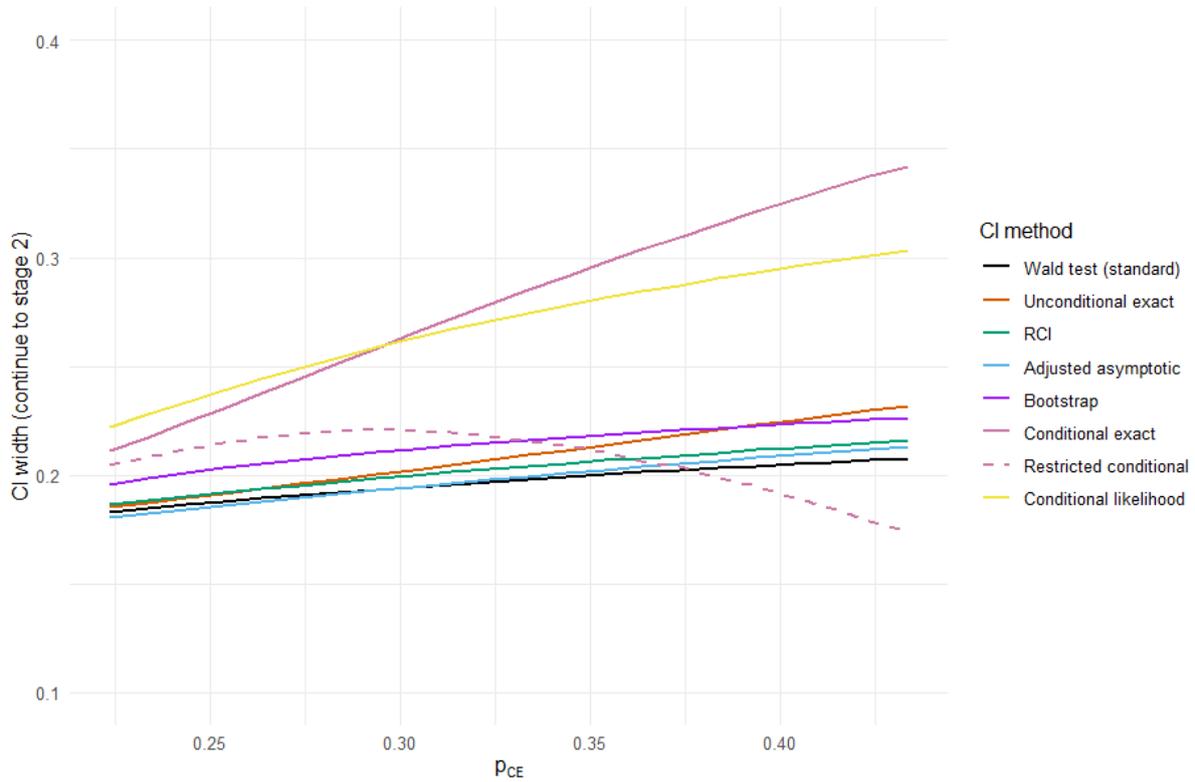

Figure 9: *CI width conditional on continuing to stage 2 as the value of $p_{CE}$ varies from $\hat{p}_{CE} - 0.07 \approx 0.224$ to $\hat{p}_{CE} + 0.14 \approx 0.434$. The value of $p_p = 21/134 \approx 0.157$.*

**Summary**

Our aim with the simulation study was not primarily to suggest that one method is the 'best' overall and should be used in the MUSEC trial context, as which method is 'best' very much depends on the trial aims and relative importance of relevant metrics of interest (including coverage, consistency and CI width), see the following Sections. However, we can draw the following general observations:

- There is often (but not always) the following trade-off in metrics: a high(er) coverage implies high(er) mean CI width, and vice-versa. The trade-off between coverage and consistency is much less clear.
- While unconditional CIs can have good performance unconditionally, the conditional performance (especially in terms of coverage) may be poor.
- To guarantee conditional performance in terms of coverage, conditional CIs are a must. However, this can come at the price of (very) wide CIs.
- For conditional CIs, as has been discussed in the literature, the restricted conditional exact CI is to be preferred to the conditional exact CI, and the penalised likelihood CI is to be preferred to the conditional likelihood CI (particularly for trials that stop early at stage 1).

- Some of the undercoverage and inconsistency is driven by the quality of the (asymptotic) normal approximation used. However, this is not unique to ADs, with such issues also seen for CIs for binary endpoints in fixed trial designs[17].
- The only CI methods that guarantee consistency in all cases in this trial context are the RCI, and the penalised likelihood CI (when stopping at stage 1).
- The simulations highlight the importance of looking at a wide(r) region of the parameter space, as some CI methods may perform well in some parts of the parameter space and not in others.
- More 'extreme' conditional results can be seen as the probability of stopping at that stage gets closer to zero or one.

# 5. Guidance: best practice for CIs in ADs

In this Section, we give guidance on the choice and reporting of CIs for ADs. This builds on the relevant parts of the FDA guidance for ADs[11] and the ACE guidance[2,3], and closely follows the guidance for best practices for point estimation in ADs given by Robertson et al. (2023a,b)[9,10]. The choice of CIs should be considered throughout an adaptive trial, from the planning stage to the final reporting and interpretation of the results. Indeed, the design and analysis of an adaptive trial are closely linked, and should ideally go hand-in-hand. In what follows, our main focus is on the confirmatory setting where analyses are fully pre-specified, but some of the principles can also apply to more exploratory settings (e.g. the CONSORT Dose-Finding Extension guidelines mention the reporting of confidence intervals[26]).

## 5.1 Planning stage

The context, aims and design of an adaptive trial should all inform the analysis strategy used, including the choice of CIs. These decisions are not only for trial statisticians, but should also be discussed with other trial stakeholders to ensure consistency with what they want to achieve. Firstly, it is necessary to decide on what exactly is to be estimated (that is, the estimands of interest[24]). Secondly, the desired characteristics of potential CIs should be decided. Some key considerations are as follows:

- *Conditional versus unconditional perspective*: The choice of whether to look at the conditional or unconditional properties of CIs will depend on the trial design. For example, in a drop-the-losers trial where only a single treatment is selected for analysis in the final stage, a conditional perspective reflects the primary interest being in estimating the treatment effect of the selected treatment. On the other hand, for group sequential trials, the unconditional perspective is recognised as being an important consideration[10]. As seen in the simulation study in Section 4, the conditional properties of unconditional CIs can be poor, while the conditional CIs can have (much) larger widths than the unconditional CIs. A general framework of viewing the question of conditional versus unconditional inference is provided by Marschner (2021)[27]. Rather than advocating for or against unconditional inference over conditional inference in general, the framework allows for the exploration of the extent to which conditional bias is likely to be present within a given sample (using meta-analysis techniques).

- *Trade-off in metrics*: As highlighted at the end of Section 4, typically there will be a trade-off between the coverage and the width of CI methods. In terms of interpretability/communication, consistency with the test decision is key. Depending on the context and aims of the trial, different relative importance may be given to the other criteria. For example, in a phase II trial where a precise estimate of the treatment effect is needed to inform a follow-up confirmatory study, the CI width may be of greater concern, whereas in a definitive phase III trial more emphasis may be placed on having the correct coverage to satisfy regulatory concerns. We are not aware of clear proposals in the literature on how to combine these different metrics into one overall criterion (as opposed to, say, the mean squared error (MSE) for point estimators).

- *Link with point estimation*: As highlighted in Sections 3 and 4, all of the CI methods (apart from the RCI) have a natural associated point estimator. Hence the considerations and guidance around choice of point estimators given in Robertson et al. (2023a,b)[9,10] can also play a role in the choice of CI method. Ideally, these choices should go hand-in-hand to avoid the (rare) situations where the chosen point estimator lies outside the chosen CI.

In trials with multiple outcomes (e.g., primary and secondary outcomes), there may be different criteria and hence CIs needed for different outcomes. As well, in some trial settings such as multi-arm trials where more than one arm can reach the final stage, the CI of each arm's comparison with control could be considered separately, but there may also be interest in calculating e.g., the simultaneous coverage across all arms that are selected. Once the criteria for assessing CIs have been decided, the next step is to find potential CIs that can be used for the trial design in question. Part I of this paper series is a starting point to find the relevant methodological literature (and code for implementation).

For certain (more common) types of ADs, such as GSDs, a review of the literature may be sufficient to compare the different types of adjusted CIs. Otherwise, we would recommend conducting simulations to explore the properties of potential CIs given the AD. It is important to assess the CIs across a range of plausible parameter values and design scenarios, taking into account factors such as the probability of early stopping. The simulation-based approach can also be used when there are no proposed alternatives to the standard CI for the trial design under consideration. Even in this setting, we would still encourage an exploration of the properties of the standard CI. If there is undercoverage or inconsistencies with the hypothesis test decision (for example), then this can impact how the results of the trial are reported (see Section 5.4). Exploring a bootstrap approach as an alternative to the standard CI may be an option in such a scenario.

## 5.2 Pre-specification of analyses

The statistical analysis plan (SAP) and health economic analysis plan (HEAP) should include a description of the CIs that are planned to be used when reporting the results of the trial, and a justification of the choice of CIs based on the investigations conducted during the planning stage. This reflects FDA guidance, which states that there should be "prespecification of the statistical methods that will be used to […] estimate treatment effects…"[11]. The trial statistician and health economist should work together to develop plans that are complementary to both their analyses.

In settings where multiple adjusted CIs are available and are of interest, one CI should be designated the 'primary' CI for the final reporting of results, with the others included as sensitivity or supplementary analyses (depending on the estimand of interest). This is to aid clarity in the interpretation of the trial results, and to avoid 'cherry-picking' the most favourable CI after observing the trial results. We have avoided making general recommendations on which CI method to use in practice because this depends on the context and goals of the trial, as well as the type of AD in question. In addition, given that CIs for ADs is an ongoing research area, there is a risk that any such recommendations may become outdated.

## 5.3 Data Monitoring Committees (DMCs)

When presenting interim results to DMCs, the choice of CIs should also be considered. For example, for GSDs the RCI has been suggested as a useful data monitoring tool[28].

## 5.4 Reporting results for a completed trial

When reporting results for a completed adaptive trial, there should be a clear description of the "statistical methods used to estimate measures of treatment effects"[2]. Hence, it should be made clear what CI method is used, along with any underlying assumptions made in their calculation (for example, being conditional on the observed stopping time). These discussions would naturally link back to the planning stage literature review and/or simulations (which could potentially be updated in light of the trial results and any unplanned adaptations that took place). For example, if the potential undercoverage of the standard CI is likely to be negligible, this would be a reassuring statement to make. On the other hand, in a setting where no adjusted CIs currently exist in the literature and there is the potential for undercoverage or any other performance issue of the standard CI, a statement flagging up this potential concern would allow appropriate caution to be taken when using the CI to inform clinical or policy decisions, future studies or meta-analyses. As discussed in Section 5.2, it should be specified in advance (i.e., in the SAP for a confirmatory study) which CI will be used for the primary analysis and which (if any) CI(s) will be used as a sensitivity analysis.

# 6. Discussion

There is a growing body of methodological literature proposing various adjusted CI methods for a wide variety of ADs, with GSDs in particular having a large number of different options, as illustrated in our case study and simulation results. However, in our experience there is at best limited uptake of adjusted CIs in practice, with many adaptive trials continuing to only report the standard CI.

It is our hope that this paper series will encourage the increased use and reporting of adjusted CIs in practice for ADs wherever possible. As described in our guidance in Section 5, estimation issues should be considered in the design stage of an adaptive trial. The estimation strategy should take the design of the trial into account, which motivates the use of adjusted CIs. In terms of trial reporting, statements about the potential undercoverage (for example) of the reported CIs can indicate where more care is needed in interpretation of the results and the use of these CIs for further research.

For future research, it would be helpful to have stronger guidance on how to choose CIs in practice for a given AD type, especially in terms of proposals around how to appropriately combine different metrics/performance measures of interest. There is also the need for the further development of user-friendly software and code for calculation of adjusted CIs in practice and to aid in simulations.

# Acknowledgements

T Jaki and DS Robertson received funding from the UK Medical Research Council (MC_UU_00002/14 and MC_UU_0040_03). B Choodari-Oskooei was supported by the MRC grant (MC_UU_00004_09 and MC_UU_12023_29). The Centre for Trials Research receives infrastructure funding from Health and Care Research Wales and Cancer Research UK.

# Data Availability Statement

All of the data that support the findings of this study are available within the paper and data files. For the purpose of open access, the author has applied a Creative Commons Attribution (CC BY) licence to any Author Accepted Manuscript version arising.

# Appendix

## A.1 Case study: Group sequential design

**Definition of the information at stages 1 and 2**

At stage $k$ ($k = 1,2$), let $\widetilde{p}_k$ denote the pooled estimate of the mean overall success probability, i.e., the total number of observed successes divided by the total number of subjects. Then the observed information $I_k$ is given by

$$I_k = \frac{1}{\widetilde{p}_k(1-\widetilde{p}_k)(1/n_{0k} + 1/n_{CEk})}$$

where $n_{0k}$ and $n_{CEk}$ are the number of subjects on the placebo and CE arms, respectively, at stage $k$.

**Definition of the conditional density of $\hat{\theta}$**

The conditional density of $\hat{\theta}$, conditional on continuing to stage 2, is given by the following expression:

$$f(\hat{\theta}|\theta, T=2) = \frac{1 - \Phi(\frac{e_1/\sqrt{I_1}-\hat{\theta}}{1/I_1-1/I_2})}{1 - \Phi(e_1 - \theta\sqrt{I_1})} \frac{\exp\left[-\frac{I_2}{2}(\hat{\theta}-\theta)^2\right]}{\sqrt{2\pi/I_2}}$$

**Unconditional parametric bootstrap procedure**

1) Given the end-of-trial response rate estimates $\hat{p}_{CE}$ and $\hat{p}_P$, generate $B$ bootstrap stage 1 samples $S_{1,CE}^{(1)}, \ldots, S_{1,CE}^{(B)}$ and $S_{1,P}^{(1)}, \ldots, S_{1,P}^{(B)}$, where $S_{1,CE}^{(b)} \sim \text{Bin}(n_{1,CE}, \hat{p}_{CE})$ and $S_{1,P}^{(b)} \sim \text{Bin}(n_{1,P}, \hat{p}_P)$ represent the bootstrap number of successes on the CE and placebo arm, respectively, at the end of stage 1 for b = 1, …, B. Here $n_{1,CE}$ and $n_{1,P}$ are the stage 1 sample sizes for the CE and placebo arm, respectively.

1) For $b = 1, \ldots, B$ calculate the bootstrap standardised stage 1 test statistic $Z_1^{(b)}$ from the bootstrap values $S_{1,CE}^{(b)}$ and $S_{1,P}^{(b)}$.

    a) If $Z_1^{(b)} > e_1$ then the bootstrap MLE $\hat{\theta}^{(b)}$ is set equal to the stage 1 MLE i.e.,
    $$\hat{\theta}^{(b)} = S_{1,CE}^{(b)}/n_{1,CE} - S_{1,P}^{(b)}/n_{1,P}$$

    b) Otherwise, generate a bootstrap stage 2 sample $S_{2,CE}^{(b)}$ and $S_{2,P}^{(b)}$, where $S_{2,CE}^{(b)} \sim \text{Bin}(n_{CE} - n_{1,CE}, \hat{p}_{CE})$ and $S_{2,P}^{(b)} \sim \text{Bin}(n_p - n_{1,P}, \hat{p}_P)$ represent the

bootstrap number of successes on the CE and placebo arm, respectively, for stage 2 only. Then the bootstrap MLE $\hat{\theta}^{(b)}$ is set equal to the overall MLE i.e.,

$$\hat{\theta}^{(b)} = (S_{1,CE}^{(b)} + S_{2,CE}^{(b)})/n_{CE} - (S_{1,P}^{(b)} + S_{2,P}^{(b)})/n_P$$

2) The bootstrap CI is then given by $(q_{\alpha/2}, q_{1-\alpha/2})$ where $q_{\alpha/2}$ and $q_{1-\alpha/2}$ are the $\alpha$ and $(1-\alpha/2)$ quantiles, respectively, of the set $\hat{\theta}^{(1)}, \ldots, \hat{\theta}^{(B)}$.

**Bootstrap procedure for the conditional likelihood CI (conditional on continuing to stage 2)**

1) Set $b = 1$.

2) Given the end-of-trial response rate estimates $\hat{p}_{CE}$ and $\hat{p}_P$, generate bootstrap stage 1 sample $S_{1,CE}^{(b)}$ and $S_{1,P}^{(b)}$, where $S_{1,CE}^{(b)} \sim Bin(n_{1,CE}, \hat{p}_{CE})$ and $S_{1,P}^{(b)} \sim Bin(n_{1,P}, \hat{p}_P)$.

3) Calculate the bootstrap standardised stage 1 test statistic $Z_1^{(b)}$ from the bootstrap values $S_{1,CE}^{(b)}$ and $S_{1,P}^{(b)}$.

    a) If $Z_1^{(b)} > e_1$, go back to step 2.

    b) Otherwise, generate a bootstrap stage 2 sample $S_{2,CE}^{(b)}$ and $S_{2,P}^{(b)}$, where $S_{2,CE}^{(b)} \sim Bin(n_{CE} - n_{1,CE}, \hat{p}_{CE})$ and $S_{2,P}^{(b)} \sim Bin(n_p - n_{1,P}, \hat{p}_P)$. The bootstrap conditional MLE $\hat{\theta}_c^{(b)}$ is calculated following the equation above, using the bootstrap values $S_{1,CE}^{(b)}, S_{2,CE}^{(b)}, S_{1,P}^{(b)}$ and $S_{2,P}^{(b)}$. Set $b = b + 1$ and go to step 2.

4) The bootstrap CI is then given by $(q_{\alpha/2}, q_{1-\alpha/2})$ where $q_{\alpha/2}$ and $q_{1-\alpha/2}$ are the $\alpha$ and $(1-\alpha/2)$ quantiles, respectively, of the set $\hat{\theta}_c^{(1)}, \ldots, \hat{\theta}_c^{(B)}$.

**Bootstrap procedure for the conditional likelihood CI (conditional on early stopping at stage 1)**

1) Set $b = 1$.

2) Given the stage 1 success probability estimates $\hat{p}_{CE}$ and $\hat{p}_P$, generate bootstrap stage 1 sample $S_{1,CE}^{(b)}$ and $S_{1,P}^{(b)}$, where $S_{1,CE}^{(b)} \sim Bin(n_{1,CE}, \hat{p}_{CE})$ and $S_{1,P}^{(b)} \sim Bin(n_{1,P}, \hat{p}_P)$

3) Calculate the bootstrap standardised stage 1 test statistic $Z_1^{(b)}$ from the bootstrap values $S_{1,CE}^{(b)}$ and $S_{1,P}^{(b)}$.

   a) If $Z_1^{(b)} > e_1$ then the bootstrap conditional MLE $\hat{\theta}_c^{(b)}$ is calculated following the equation above, using the bootstrap values $S_{1,CE}^{(b)}$ and $S_{1,P}^{(b)}$. Set $b = b + 1$ and go to step 2.

   b) Otherwise, go back to step 2.

4) The bootstrap CI is then given by $(q_{\alpha/2}, q_{1-\alpha/2})$ where $q_{\alpha/2}$ and $q_{1-\alpha/2}$ are the $\alpha/2$ and $(1 - \alpha/2)$ quantiles, respectively, of the set $\hat{\theta}_c^{(1)}, \ldots, \hat{\theta}_c^{(B)}$.

## A.2 Additional simulation results

Simulations with the true success rates $(p_p, p_{CE})$ given by $p_p = 21/134 \approx 0.157$ and $p_{CE} = 42/143 + 0.08 \approx 0.374$. The probability of stopping early for efficacy in stage 1 is 0.761.

**Overall (unconditional results)**

| Type of CI | CI method | Coverage | Mean width (se) | Consistency | $P(L(X) > \theta)$ | $P(U(X) < \theta)$ |
|---|---|---|---|---|---|---|
| **Standard/naive** | Wald test | 0.948 | 0.227 (0.016) | 0.999 | 0.028 | 0.025 |
| **Unconditional** | Exact | 0.958 | 0.240 (0.015) | 0.999 | 0.019 | 0.023 |
|  | Repeated | 0.977 | 0.318 (0.062) | 1.000 | 0.001 | 0.022 |
|  | Adjusted asymptotic | 0.956 | 0.236 (0.019) | 0.998 | 0.019 | 0.025 |
|  | Parametric bootstrap | 0.960 | 0.218 (0.009) | 0.999 | 0.029 | 0.010 |
| **Conditional** | Exact | 0.954 | 0.456 (0.261) | 0.587 | 0.020 | 0.026 |
| | Restricted exact | 0.954 | 0.258 (0.061) | 0.997 | 0.020 | 0.026 |
| | Likelihood | 0.988 | 0.698 (0.356) | 0.351 | 0.009 | 0.003 |
| | Penalised likelihood | 0.986 | 0.308 (0.024) | 0.999 | 0.011 | 0.003 |

Table 7: Simulation results showing the performance of various CIs with $p_p = 21/134 \approx 0.157$ and $p_{CE} = 42/143 + 0.08 \approx 0.374$. There were $10^5$ trial replicates. The probability of stopping at stage 1 is 0.761.

**Conditional on stopping at stage 1**

| Type of CI | CI method | Coverage | Mean width (se) | Consistency | $P(L(X) > \theta)$ | $P(U(X) < \theta)$ |
|---|---|---|---|---|---|---|
| **Standard/naive** | Wald test | 0.964 | 0.235 (0.009) | 1.000 | 0.036 | 0.000 |
| **Unconditional** | Exact | 0.975 | 0.246 (0.009) | 1.000 | 0.025 | 0.000 |
| | Repeated | 0.998 | 0.352 (0.013) | 1.000 | 0.002 | 0.000 |
| | Adjusted asymptotic | 0.975 | 0.246 (0.009) | 1.000 | 0.025 | 0.000 |
| | Parametric bootstrap | 0.961 | 0.218 (0.009) | 1.000 | 0.039 | 0.000 |
| **Conditional** | Exact | 0.957 | 0.501 (0.283) | 0.461 | 0.017 | 0.026 |
| | Restricted exact | 0.957 | 0.275 (0.052) | 1.000 | 0.017 | 0.026 |

| | | | | | |
|---|---|---|---|---|---|
| | Likelihood | 0.988 | 0.826 (0.312) | 0.149 | 0.012 | 0.000 |
| | Penalised likelihood | 0.986 | 0.314 (0.022) | 1.000 | 0.014 | 0.000 |

*Table 8: Simulation results showing the performance of various CIs with $p_p = 21/134 \approx 0.157$ and $p_{CE} = 42/143 + 0.08 \approx 0.374$. There were $10^5$ trial replicates. The probability of stopping at stage 1 is 0.761.*